\documentclass[aps,prb,reprint,showpacs,twocolumn,longbibliography]{revtex4-1}

\usepackage{graphicx}% Include figure files
\usepackage{dcolumn}% Align table columns on decimal point
\usepackage{bm}% bold math
\usepackage{braket}
\usepackage{leftidx}
\usepackage{cancel}
\usepackage{amsmath}
\usepackage{color}
\usepackage[mathcal]{eucal}

\begin{document}

\title{Fano enhancement of SERS signal without increasing the hot spot intensity}

\author{Selen Postaci$^{1,2}$}\thanks{saatciselen@gmail.com}
\author{Bilge Can Yildiz$^{3}$} %\thanks{bilge.yildiz@metu.edu.tr}
\author{Alpan Bek$^{2, 4, 5}$} %\thanks{bek@metu.edu.tr}
\author{Mehmet Emre Tasgin$^{1}$} %\thanks{metasgin@hacettepe.edu.tr}

\address{$^{1}$Institute of Nuclear Sciences, Hacettepe University, 06800, Ankara, Turkey \\
$^{2}$Department of Physics, Middle East Technical University, 06800, Ankara, Turkey \\
$^{3}$Department of Applied Physics, Atilim University, 06836, Ankara, Turkey  \\
$^{4}$The Center for Solar Energy Research and Applications (GUNAM), Middle East Technical University, 06800, Ankara, Turkey \\ 
$^{5}$Micro and Nanotechnology Program of Graduate School of Natural and Applied Sciences, Middle East Technical University, 06800, Ankara, Turkey}

\date{\today}

\begin{abstract}
Plasmonic nanostructures enhance nonlinear response, such as surface enhanced Raman scattering (SERS), by localizing the incident field into hot spots. The localized hot spot field can be enhanced even further when linear Fano resonances (FR) take place in a double resonance scheme. However, hot spot enhancement is limited with the modification of the vibrational modes, the break-down of the molecule and the tunnelling regime. Here, we present a method which can circumvent these limitations. Our analytical model and solutions of 3D Maxwell equations show that:
enhancement due to the localized field can be multiplied by a factor of $10^2$ to $10^3$. Moreover, this can be performed without increasing the hot spot intensity which also avoids the modification of the Raman modes. Unlike linear Fano resonances, we create a path interference in the nonlinear response. We demonstrate on a single equation that enhancement takes place due to cancellation of the contributing terms in the denominator of the  SERS response.  
\end{abstract}

%\pacs{03.67.Bg, 03.67.Mn, 42.50.Dv, 42.50.Ex}
%\keywords{}

\maketitle

\section{Introduction}

%Nonlinear Effects and Confinement of Light

Metal nanoparticles (MNPs) confine incident electromagnetic field into nm-size hot spots as plasmonic oscillations. Field intensity at the hot spots can be 5 orders of magnitude larger compared to the incident one~\cite{stockman2011nanoplasmonics, wu2010quantum}. It is also reported that self-repeating cascaded materials can confine light even better compared to the gaps between MNPs~\cite{hoppener2012self, he2016near}. Intense fields give rise to appearance of nonlinear processes such as second harmonic generation (SHG), four wave mixing (FWM), and surface enhanced Raman scattering (SERS)~\cite{kauranen2012nonlinear, hua2014nature, wang2008manipulation}. Actually, enhancement is squared since the field of the converted frequency is also localized~\cite{ding2016nanostructure}. In an efficient conversion, nonlinear process takes place between plasmonic excitations of different frequencies due to the localization~\cite{grosse2012nonlinear, turkpence2014engineering, chu2010double}. Recent studies show that MNPs with plasmon resonances at both excitation and Stokes frequencies (double resonance) provide better enhancement factors for Raman intensities~\cite{chu2010double, Mueller2016, Jorio2017}.

Hot spots also provide enhanced light-matter interaction. When a quantum emitter (QE) is placed into a hot spot, localized plasmon field interacts strongly with the QE. Small decay rate of the QE creates Fano resonances, a dip in the plasmonic spectrum~\cite{luk2010fano, limonov2017fano}. In this process, the localized plasmon field provides the weak hybridization. Fano resonance also appears when excited plasmon mode couples to a long-live dark plasmon mode~\cite{luk2010fano, limonov2017fano, tassin2009low, liu2009plasmonic}.

Fano resonances can extend the lifetime of plasmon excitations~\cite{Sadeghi2015, tacsgin2013metal, ElKabbash2017, bilgethesis} which makes the operation of coherent plasmon emission (spaser) possible~\cite{noginov2009demonstration}. They also lead to further enhancement of the localized hot spot field~\cite{Stockman2010}. This extra enhancement in the hot spot field is cleverly adopted for the enhancement of the nonlinear response in FWM~\cite{Zhang2013a} and SERS~\cite{ye2012plasmonic, Zhang2014, he2016near}. Similar to double resonance scheme~\cite{chu2010double}, both the excited and Stokes shifted frequencies are aligned with two Fano resonances~\cite{ye2012plasmonic, Zhang2014, he2016near}. The double Fano resonance scheme provides much stronger enhancement in the SERS signal. Fano resonances are also shown to provide control over other nonlinear processes such as SHG~\cite{tasgin2016fluorescence, butet2012nonlinear}, third harmonic generation~\cite{shorokhov2016multifold}, and FWM~\cite{singh2016enhancement, paspalakis2014strongly}.

SERS is a very useful imaging technique. It provides information about the chemical composition of newly synthesized molecules by determining the existing bond types. Single-molecule detection via SERS is studied in many fields of science, including chemistry~\cite{zhang2013chemical}, nanobiology~\cite{kneipp2008sers}, tumor targeting and cancer applications~\cite{qian2008vivo}. Even more, mapping of inner structure and surface configuration of a single molecule is achieved recently~\cite{zhang2013chemical} using a double resonance scheme~\cite{chu2010double}. Such an imaging requires very intense fields at the nm-size hot spots. When the imaging tip gets closer to the metal surface, the intensity at the hot spot --where the molecule lies-- increases. If the hot spot intensity is increased further, e.g. via a double Fano resonance scheme~\cite{ye2012plasmonic, Zhang2014, he2016near}, fragile molecules can be damaged~\cite{yang2013ultraviolet, Schaffer2001}. It is also experimented that vibrational modes of a Raman-imaged nanostructure (i.e. a carbon nanotube) can be modified due to the close spacing of the tip~\cite{CNTRaman2006}. Additionally, electron tunnelling can limit the intensity enhancement in the gaps~\cite{Zhu2014}.    

In this manuscript, we study the SERS signal from a double resonance system. A Raman reporter molecule is placed close to the gap of a MNP dimer, see Fig.~\ref{fig1}. We additionally place an auxiliary QE (e.g. a molecule or a nitrogen vacancy centre) to the other side of the gap. 3D solutions of Maxwell equations show that SERS can be enhanced by a factor of $10^3$ without increasing the field intensities at the excited and the Stokes-shifted hot spots. This enhancement multiplies the enhancement due to localization. On a basic analytical model, we demonstrate the underlying reason for the enhancement. Coupling of the auxiliary QE with the Stokes-shifted plasmon mode modifies frequency conversion paths dramatically. It yields a cancellation in the denominator of the SERS response, i.e. Eq.~(\ref{R_scattering}). 3D simulations show that enhancement predicted by the analytical model, also appears in the presence of retardation effects. 

The presented phenomenon can be adopted to further increase the efficiency of SERS imaging for systems which are already operating in the break-down or tunnelling regimes. Better signal intensities with larger tip-surface spacing or with smaller laser intensities can be achieved by avoiding modifications in the Raman vibrational modes.

Our problem setting, which involves configuration of a MNP dimer coupled to a Raman reporter molecule and an auxiliary QE, can be implemented controllably using several nanotechnological methods such as e-beam lithography~\cite{santhosh2016vacuum,hentschel2016linear} or DNA based biomolecular recognition~\cite{liu2006nanoplasmonic, barrow2012surface} that provide ultimate nanoscale spatial control~\cite{supp}. One can also conduct an experiment based on the stochastic distributions of many molecules~\cite{tasgin2016fluorescence}. A practical implementation would be the following. A gold coated AFM tip decorated (can also be considered as contamination) with carefully chosen auxiliary molecules (QEs) as shown in Fig.~\ref{fig2}, using a technique reminiscent of dip-pen lithography, will produce more intense SERS signal without increasing near-field intensity. In spasers~\cite{noginov2009demonstration}, where MNPs are surrounded by molecules, linear Fano resonance increases the plasmon lifetime and fluorescence intensity of the molecules~\cite{hoppener2012self}. Fano resonances can also be adopted in an all-plasmonic setting~\cite{hsiao2016enhancement,thyagarajan2013augmenting}. 

In the following, we first present the basic analytical model from which we anticipate the presence of the enhancement. We introduce the effective Hamiltonian for a double resonance SERS system coupled with an auxiliary QE. We obtain the equations of motion (EOM) using Heisenberg equations. We manage to obtain a simple expression for the steady-state of the Stokes field amplitude, Eq.~(\ref{R_scattering}). On this expression we explain why such an enhancement should emerge. Next, we perform simulations of the exact solutions of the 3D Maxwell equations in order to test the retardation effects. In this case, the spectrum for which enhancement appears narrows down compared to the analytical result. Even so, an enhancement of 3 orders of magnitude on top of the localization can be observed.

\section{Hamiltonian and Equation of Motion}

\begin{figure}
\centering
\includegraphics[trim=6cm 0cm 7cm 0cm, clip, width=0.5\textwidth]{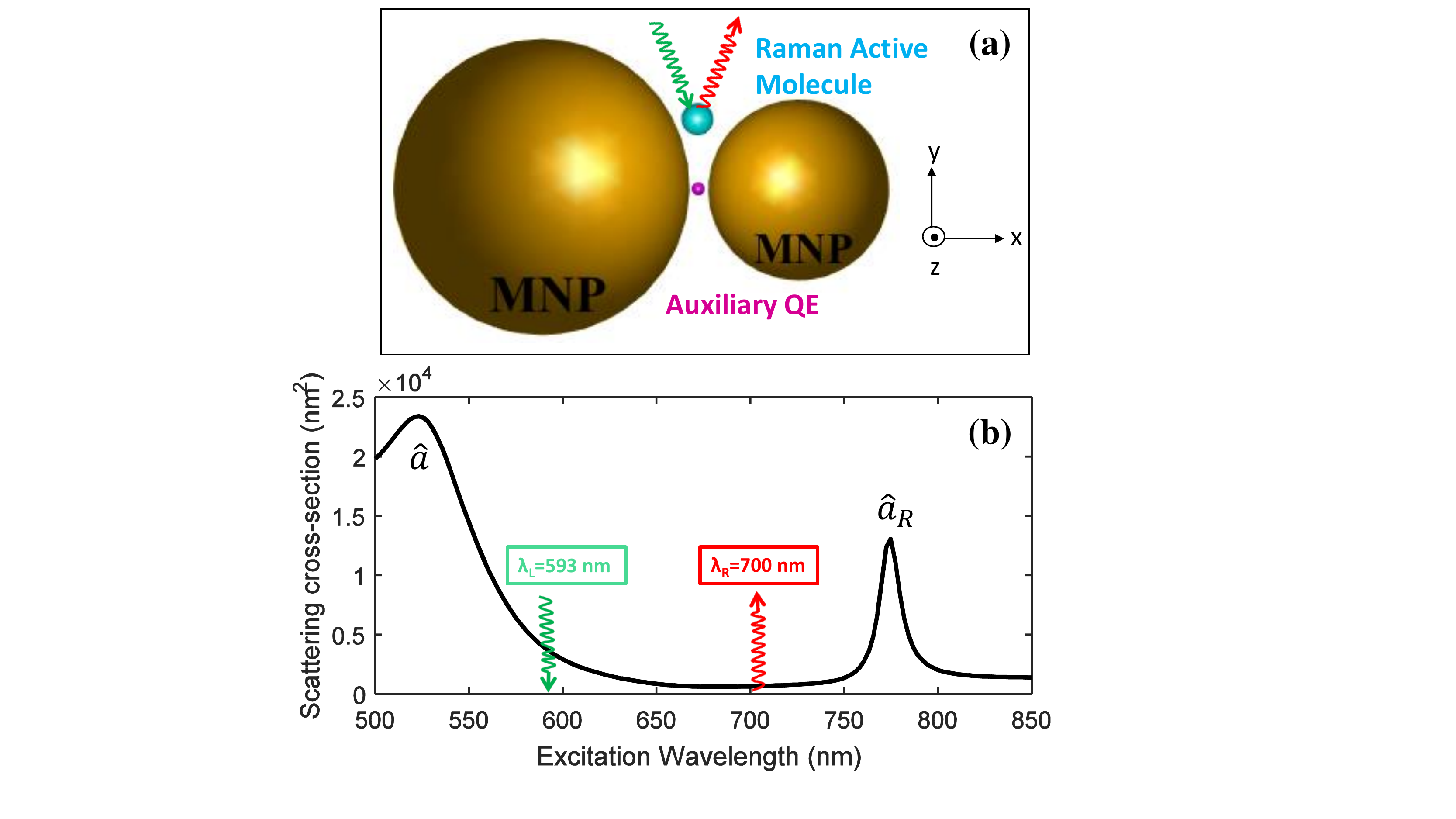}
\caption{(a) The setup we use in the 3D solutions of Maxwell equations. A gold MNP dimer, of radii 90 nm and 55 nm, creates a hot spot in the 4 nm gap. A sphere of 4 nm radius (blue), representing a Raman reporter, is placed close to the hot spot for producing the SERS signal. We place an auxiliary QE (purple) also at the hot spot of the dimer, for enhanced interaction with the $\hat{a}_{\rm R}$ plasmon mode. We move it along the z-direction when we desire to decrease the plasmon-auxiliary QE coupling, $f$. (b) Linear response of the dimer shows two plasmon peaks at $\Lambda$=530 nm and $\Lambda_{\rm R}$=780 nm. System is excited by a $\lambda_{\rm L}$=593 nm laser and a Stokes shifted signal emerges at $\lambda_{\rm R}$=700 nm. $\lambda_{\rm L}$ and $\lambda_{\rm R}$ overlap with $\Lambda$ and $\Lambda_{\rm R}$, respectively. The $\lambda_{eg}$ of the molecule is chosen to couple with the $\hat{a}_{\rm R}$-mode, see Fig.~\ref{fig3}a. We use parameters similar to (b) in producing an accompanying simulation within our simple model Eq.~(\ref{H_R}). Experimental data of gold (dimer) and a Lorentzian dielectric function (auxiliary QE), in MNPBEM~\cite{hohenester2012mnpbem}, are used for the 3D simulations. }
\label{fig1}
\end{figure}

\begin{figure}
\centering
\includegraphics[width=0.45\textwidth]{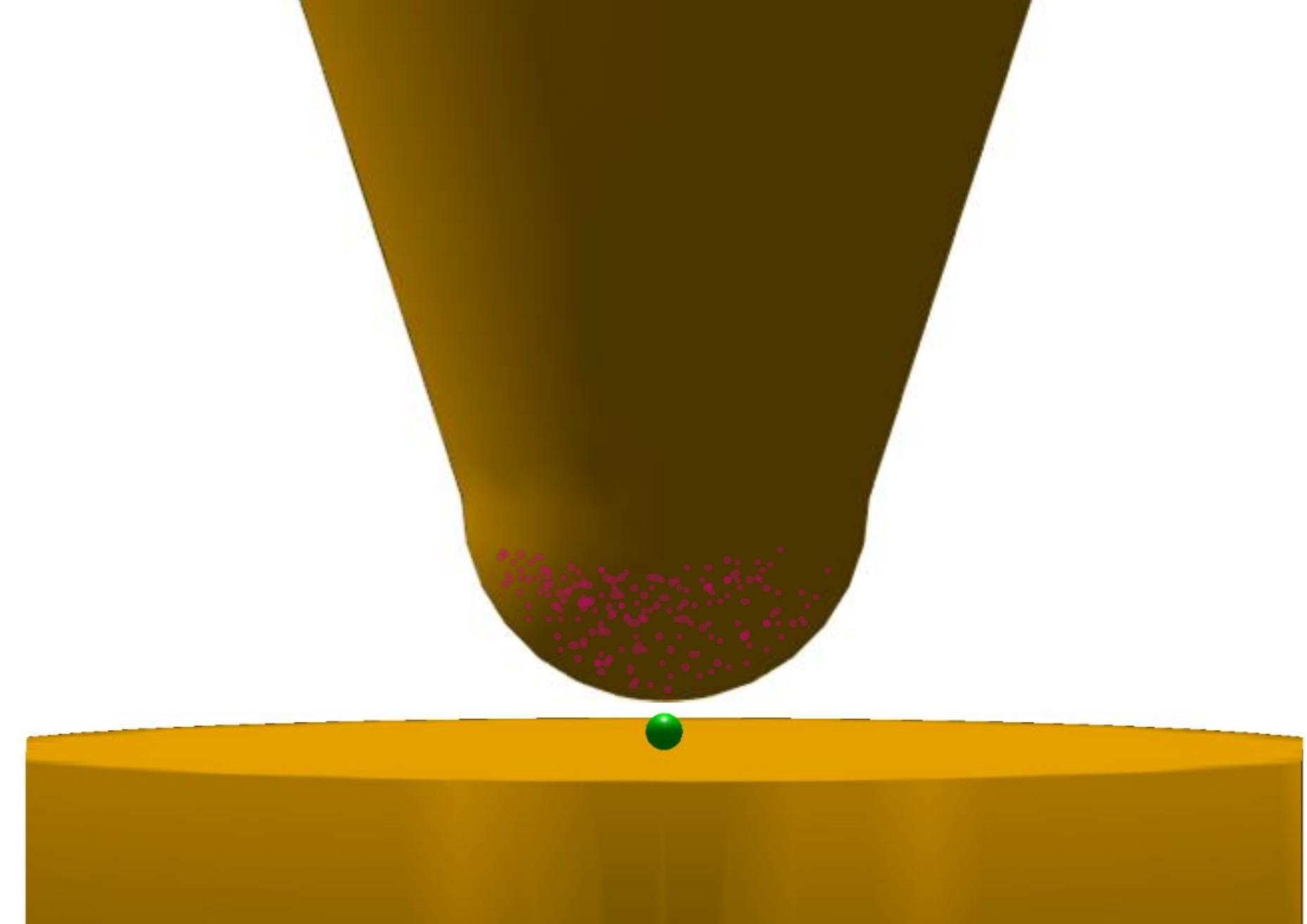}%[trim=6cm 3cm 9cm 9cm, clip, width=0.5\textwidth]{fig2}
\caption{A practical implementation of the extra enhancement. A gold coated AFM tip is decorated (can as well be named as contamination) by the auxiliary QEs (purple). The auxiliary QEs, with a level spacing not close to either the exciting or the SERS frequency, do not cause an alteration of the hot-spot field intensity at the location of the Raman reporter molecule (blue), yet they give rise to further enhancement of the SERS from reporter molecule due to Fano effect.}
\label{fig2}
\end{figure}

We consider a double resonance scheme with two plasmon bands, $\hat{a}$ and $\hat{a}_{\rm R}$, with resonances $\Lambda=c/\Omega$=532 nm  and $\Lambda_{\rm R}=c/\Omega_{\rm R}$=780 nm respectively, see Fig.~\ref{fig1}. A strong incident laser field, $\lambda_{\rm L}=c/\omega$=593 nm, excites the plasmon polaritons in the $\hat{a}$-mode. The substantial overlap between the hot spots of the two modes, $\hat{a}$ and $\hat{a}_{\rm R}$, and the Raman reporter molecule yields a significant overlap integral $\chi$ for the Stokes Raman process. Hence, a plasmon in the excited $\hat{a}$ mode generates a Stokes shifted plasmon polariton with $c/\omega_{\rm R}=\lambda_{\rm R}$=700 nm in the lower energy mode $\hat{a}_{\rm R}$~~\cite{Jorio2017,Mueller2016,chu2010double}.

When an auxiliary QE is inserted in the system, it also interacts strongly with the hot spot of $\hat{a}_{\rm R}$-mode, into which the nonlinear conversion takes place. Level spacing of the QE, $\omega_{eg}$, is chosen about $\Omega_{\rm R}$. 

Hamiltonian for such a system, including the Raman conversion, can be written as the sum of the terms $\hat{H}_{0}+\hat{H}_{\rm QE}+\hat{H}_{\rm L}+\hat{H}_{\rm int}+\hat{H}_{\rm R}$, with            
\begin{align}
\hat{H}_{0} &= \hbar\Omega\hat{a}^{\dagger}\hat{a}+\hbar\Omega_{\rm R}\hat{a}_{\rm R}^{\dagger}\hat{a}_{\rm R}+\hbar\Omega_{\rm ph}\hat{a}_{\rm ph}^{\dagger}\hat{a}_{\rm ph} \nonumber \\
\hat{H}_{\rm QE} &= \hbar\omega_{eg}\ket{e}\bra{e} \nonumber \\
\hat{H}_{\rm L} &= i\hbar(\hat{a}^{\dagger}\varepsilon e^{-i\omega t}-\hat{a}\varepsilon^*e^{i\omega t}), \nonumber \\
%\hat{H}_{\rm L} &= i\hbar(\hat{a}^{\dagger}\varepsilon e^{-i\omega t}-\hat{a}\varepsilon^*e^{i\omega t}) \label{H_L}, \\
\hat{H}_{\rm int} &= \hbar(f\hat{a}_{\rm R}\ket{e}\bra{g}+f^*\hat{a}_{\rm R}^{\dagger}\ket{g}\bra{e}), \nonumber \\
\hat{H}_{\rm R} &= \hbar\chi(\hat{a}_{\rm R}^{\dagger}\hat{a}_{\rm ph}^{\dagger}\hat{a}+\hat{a}^{\dagger}\hat{a}_{\rm ph}\hat{a}_{\rm R}), \label{H_R} 
\end{align}
where $\hat{H}_{0}$ includes the energies for the driven $\hat{a}$, and Raman shifted $\hat{a}_{\rm R}$ plasmon modes as well as the molecular vibrations, $\hat{a}_{\rm ph}$. $\hat{H}_{\rm QE}$ is the energy of the auxiliary QE. $\hat{H}_{\rm L}$ is the laser pump, $\hat{H}_{\rm R}$ denotes the Raman process and $\hat{H}_{\rm int}$ is the interaction of the Stokes-shifted plasmon polaritons of $\hat{a}_{\rm R}$-mode with the auxiliary QE. $\chi$ determines the strength of the Raman process, while $\varepsilon$ denotes the power of the incident laser source. Here, we do not consider the anti-Stokes shift in the Hamiltonian to simplify our results, however, we have verified that the enhancement values and the spectroscopic behaviour of the system remain similar in such a case. The interaction strength between the auxiliary QE and $\hat{a}_{\rm R}$-mode is denoted by $f$. Coupling of the auxiliary QE to $\hat{a}$-mode is not considered due to far-off-resonance and simplicity. $\ket{g}$ and $\ket{e}$ represents the ground and excited states for the auxiliary QE. $\hat{H}_{\rm R}$ is a standard Hamiltonian for a Raman process, described, for instance, in the Refs.~\cite{Mueller2016, Jorio2017,chu2010double}. A similar form of $\hat{H}_{\rm R}$ could have also been derived~\cite{supp} from a radiation pressure like interaction~\cite{Schmidt2016, Roelli2015a}. We obtain the dynamics via Heisenberg equations, $i\hbar\dot{\hat{a}}=[\hat{a}, \hat{H}]$. We note that, since we do not consider the quantum optical effects, we are able to replace the operators with complex numbers~\cite{Premaratne2017}; $\hat{a}\rightarrow\alpha$, $\hat{a}_{\rm R}\rightarrow\alpha_{\rm R}$, $\hat{a}_{\rm ph}\rightarrow\alpha_{\rm ph}$, $\hat{\rho}_{eg}=\ket{e}\bra{g}\rightarrow\ \rho_{eg}$. We find the EOM as 
\begin{align}
\dot{\alpha}_{\rm R} &= (-i\Omega_{\rm R}-\gamma_{\rm R})\alpha_{\rm R}-i\chi\alpha_{\rm ph}^*\alpha-if^*\rho_{ge} \label{alpha_R}, \\
\dot{\alpha} &= (-i\Omega-\gamma)\alpha-i\chi\alpha_{\rm ph}\alpha_{\rm R}+\varepsilon e^{-i\omega t}, \\
\dot{\alpha}_{\rm ph} &= (-i\Omega_{\rm ph}-\gamma_{\rm ph})\alpha_{\rm ph}-i\chi\alpha_{\rm R}^*\alpha+\varepsilon_{\rm ph}e^{-i\omega_{\rm ph}t}, \\
\dot{\rho}_{eg} &= (-i\omega_{eg}-\gamma_{eg})\rho_{eg}+if\alpha_{\rm R}(\rho_{ee}-\rho_{gg}), \\
\dot{\rho}_{ee} &= -\gamma_{ee}\rho_{ee}+if^*\alpha^*_{\rm R}\rho_{eg}-if\alpha_{\rm R}\rho^*_{eg}, \label{Rho_ee}
\end{align} 
where we introduce the damping rates $\gamma$, $\gamma_{\rm R}$, $\gamma_{\rm ph}$, $\gamma_{eg}$, and $\gamma_{ee}$. We also have the constraint $\rho_{ee}+\rho_{gg}$=1. $\varepsilon_{\rm ph}$ is introduced for the vibrations, due to the finite ambient temperature~\cite{Schmidt2016, Roelli2015a}. Its actual value has no influence in the relative enhancement/suppression factors.

In the steady-state, solutions are in the form $\alpha_{\rm R}(t) = \tilde{\alpha}_{\rm R}e^{-i\omega_{\rm R} t}$, $\alpha(t) = \tilde{\alpha}e^{-i\omega t}$, $\alpha_{\rm ph}(t) = \tilde{\alpha}_{\rm ph}e^{-i\omega_{\rm ph} t}$, $\rho_{\rm eg}(t) = \tilde{\rho}_{\rm eg}e^{-i\omega_{\rm R} t}$, $\rho_{\rm ee}(t) = \tilde{\rho}_{\rm ee}$, where exponentials cancel in each equation, Eqs.~(\ref{SSAlpha_R})-(\ref{SSRho_ee}). In other words, this is the energy conservation in the long term limit. Eqs.~(\ref{alpha_R})-(\ref{Rho_ee}) become
\begin{align}
[i(\Omega_{\rm R}-\omega_{\rm R})+\gamma_{\rm R}]\tilde{\alpha}_{\rm R} &= -i\chi\tilde{\alpha}_{\rm ph}^*\tilde{\alpha}-if^*\tilde{\rho}_{\rm eg}, \label{SSAlpha_R} \\
[i(\Omega-\omega)+\gamma]\tilde{\alpha} &= -i\chi\tilde{\alpha}_{\rm ph}\tilde{\alpha}_{\rm R}+\varepsilon, \label{SSalpha} \\
[i(\Omega_{\rm ph}-\omega_{\rm ph})+\gamma_{\rm ph}]\tilde{\alpha}_{\rm ph} &= -i\chi\tilde{\alpha}_{\rm R}^*\tilde{\alpha}+\varepsilon_{\rm ph}, \label{SSAlpha_ph} \\
[i(\omega_{eg}-\omega_{\rm R})+\gamma_{eg}]\tilde{\rho}_{\rm eg}&=if\tilde{\alpha}_{\rm R}(\tilde{\rho}_{\rm ee}-\tilde{\rho}_{\rm gg}), \label{SSRho_ge} \\
\gamma_{\rm ee}\tilde{\rho}_{\rm ee}&=-if\tilde{\alpha}_{\rm R}\tilde{\rho}^*_{\rm eg}+if^*\tilde{\alpha}_{\rm R}^*\tilde{\rho}_{\rm eg}. \label{SSRho_ee}
\end{align}

We can obtain a simple expression for the Stokes-shifted plasmon amplitude (SERS signal) by using Eqs.~(\ref{SSAlpha_R}) and (\ref{SSAlpha_ph}) 
\begin{align}
\resizebox{0.88\hsize}{!}{$
\tilde{\alpha}_{\rm R}=\frac{-i\chi\varepsilon_{\rm ph}^*}{\beta_{\rm ph}^*\left([i(\Omega_{\rm R}-\omega_{\rm R})+\gamma_{\rm R}]-\frac{|f|^2y}{[i(\omega_{eg}-\omega_{\rm R})+\gamma_{eg}]}\right)-|\chi|^2|\tilde{\alpha}|^2}\tilde{\alpha},
\label{R_scattering}
$}
\end{align}
%\end{widetext}
%\vspace{\columnsep}
%\twocolumngrid
where $\beta_{\rm ph}=[i(\Omega_{\rm ph}-\omega_{\rm ph})+\gamma_{\rm ph}]$. Here y=$\rho_{ee}-\rho_{gg}$ is the population inversion for the auxiliary QE. $|\chi|^2|\tilde{\alpha}|^2$ term is small compared to other ones in the denominator and hence, can be neglected.

We use Eq.~(\ref{R_scattering}) merely to anticipate the enhancement/suppression effects. All the presented results are obtained by numerical time evolution of Eqs.~(\ref{alpha_R})-(\ref{Rho_ee}).

{\bf  \small Enhancement}. A quick examination of the denominator of Eq.~(\ref{R_scattering}) reveals that for the proper choice of $\omega_{eg}$, nonresonant term ($\Omega_{\rm R}$-$\omega_{\rm R}$) in the denominator can be cancelled with the term containing $f$, the MNP-QE coupling. This condition is    
\begin{align}
\omega_{eg}^*=\omega_{\rm R}+\frac{|f|^2y}{2(\Omega_{\rm R}-\omega_{\rm R})}-\sqrt{\frac{|f|^4|y|^2}{4(\Omega_{\rm R}-\omega_{\rm R})^2}-\gamma_{\rm eg}^2}. \label{weg_enhancement}
\end{align}

This choice for the level spacing enables us to minimize the denominator, consequently enhancing the Raman signal amplitude. This type of enhancement does not necessitate an arrangement in the inner structure of plasmon modes.

To examine the dependence of the enhancement with respect to the level spacing ($\lambda_{eg}=c/\omega_{eg}$), we time evolve the EOM~(\ref{alpha_R})-(\ref{Rho_ee}). The parameters are chosen as $\gamma$=0.01$\omega$, $\gamma_{\rm R}$=0.005$\omega$, $\gamma_{\rm ph}$=0.001$\omega$. Nevertheless, one can realize that $\Omega_{ph}$ and $\gamma_{ph}$ play no role in the cancellation of the denominator in Eq.~(\ref{R_scattering}). The damping rate (spectral width) of the auxiliary QE is taken to be $\gamma_{eg}$=$10^{-5}\omega$. Here, the frequency of the incident light ($\omega$) is related to $\lambda_{\rm L}$ as $\omega=c/\lambda_{\rm L}$=593 nm. $\chi$ is assumed a small value $10^{-5}\omega$, where it is verified that the value of $\chi$ does not affect the enhancement factors, and $\varepsilon=0.1\omega$. $f$ is also varied in order to explore the effect of the coupling in the MNP-QE system. The enhancement factor is calculated with respect to the $|\alpha_{\rm R}|^2$ intensity for $f$=0.

\begin{figure}[h!]
\centering
\includegraphics[trim=8cm 1.2cm 0cm 0cm, clip, width=0.63\textwidth]{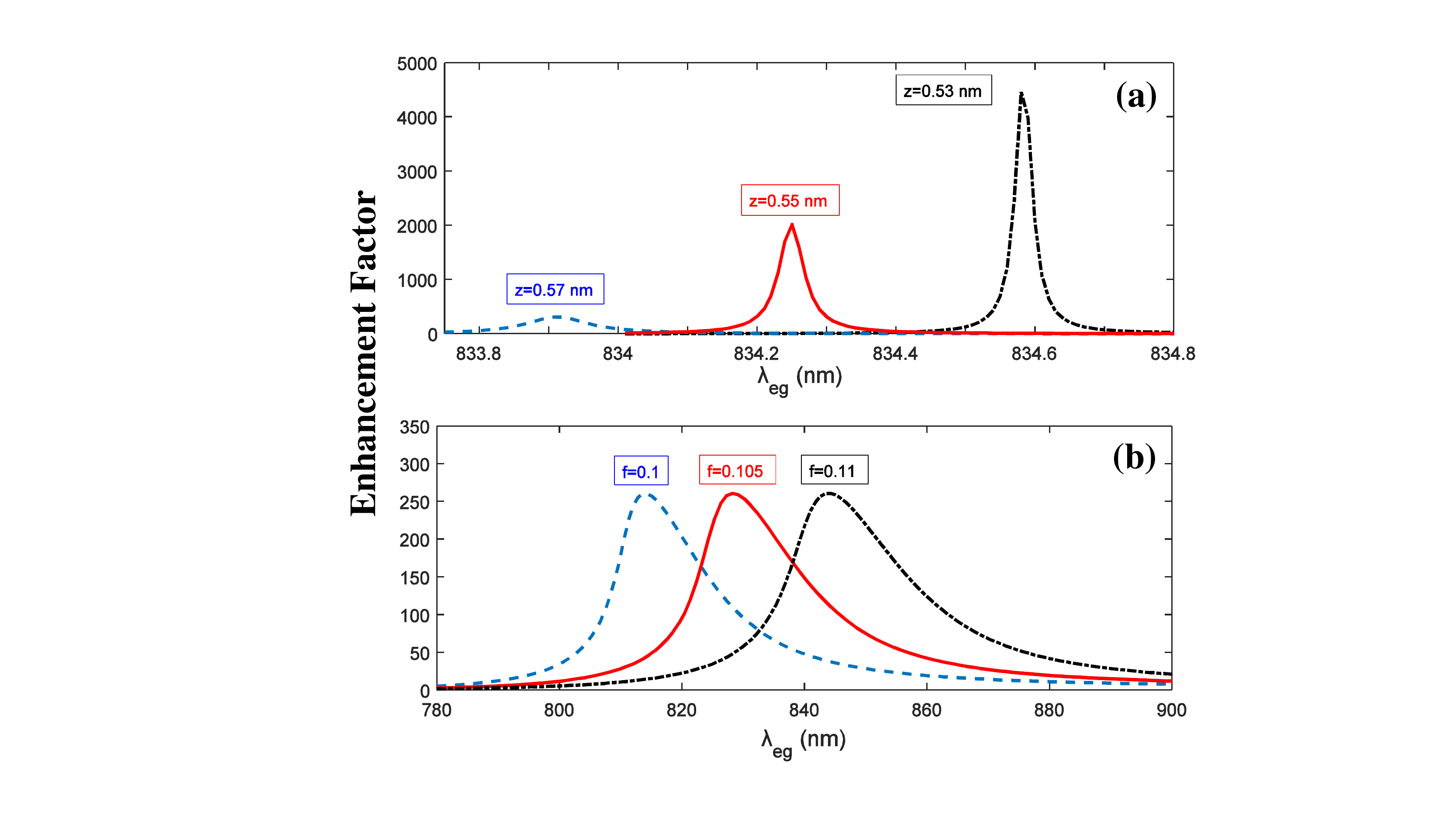}
\caption{Enhancement of the SERS signal with the presence of the auxiliary QE (purple) in Fig.~\ref{fig1}. We plot the enhancement factors with respect to the level spacing ($\lambda_{eg}=c/\omega_{eg}$) of the auxiliary QE interacting with the $\hat{a}_{\rm R}$ plasmon mode, (Fig.~\ref{fig1}b). Interaction of the auxiliary QE decreases with larger separations (z) to the hot spot. (a) Solutions of the 3D Maxwell equations for the system depicted in Fig.~\ref{fig1}. Enhancement factors, multiplying localization effects, for SERS are calculated for three different positions of the auxiliary QE, z=0.53, 0.55 and 0.57 nm. At z=0.0 nm, we observe 2 orders of enhancement factor. The position of optimum $\lambda_{eg}^*=c/\omega_{eg}^*$, where maximum enhancement factor appears, shifts to larger wavelengths. Optimum $\lambda_{eg}^* \approx$834 nm, is far away from the Stokes signal (700 nm). An enhancement due to a linear Fano resonance would yield maximum enhancement at $\lambda_{eg}\approx$ 700 nm~\cite{Stockman2010}. Enhancement originates from the interference in the conversion paths as predicted by Eqs.~(\ref{R_scattering})-(\ref{weg_enhancement}). Intensities of both hot spots are unchanged. (b) Simulation of the enhancement factor using Eqs.~(\ref{alpha_R})-(\ref{Rho_ee}) with the parameters similar to Fig.~\ref{fig1}b. Simulations are for the presence of an auxiliary QE. We observe the similar behaviour in the position of optimum $\lambda_{eg}^*$, even though several complications involve in 3D simulations (a). }
\label{fig3}
\end{figure}

The results are depicted in Fig.~\ref{fig3}(b), where enhancement factors of $\approx$300 are observed. As suggested by Eq.~(\ref{weg_enhancement}), $\Omega_{\rm R}<\omega_{\rm R}$, cancellation (enhancement) takes place for longer wavelengths as MNP-QE coupling, $f$, increases. The spectral position of $\lambda_{eg}^*=c/\omega_{eg}^*$ also justifies our assumption for off-resonant $\hat{a}$-QE coupling. If Eq.~(\ref{R_scattering}) is examined, it can be realized that the amount of enhancement can be increased by introducing more interference paths via additional QEs~\cite{singh2016enhancement} or additional plasmon conversion modes. 

Eq.~(\ref{R_scattering}) is a single and simple equation which enables us to predict possible interference effects without including the complications emerging in 3D simulations. Before moving forward, we underline that our aim is to present a simple understanding for the enhancement process, without getting lost in details.

Linear Fano resonances, commonly referred in the literature, appear if one of the two coupled oscillators has longer lifetime~\cite{luk2010fano,garrido2002classical,tassin2009low,liu2009plasmonic}. Here, interference of the nonlinear frequency conversion paths demonstrates us an interesting incident. Even when the spectral width (damping rate) of the auxiliary object is equal to the damping rate of the MNP hot spot, 25 times enhancement can emerge due to cancellation in the denominator of Eq.~(\ref{R_scattering}). The presented enhancement factor is obtained for a plasmon mode of fair quality $\gamma$=0.01$\omega$. When a higher quality MNP~\cite{minhigh2009, west2010searching} is used available enhancement factor grows up.

\section{3-Dimensional Simulations}

 We also perform simulations with the exact solutions of 3D Maxwell equations and use the setup in Fig.~\ref{fig1}. We note in advance that we do not aim a one to one comparison between the analytical solutions and the 3D simulations. We aim to observe if the retardation effects wipe-out the enhancement phenomenon predicted by our basic analytical model. Making a one to one comparison between the theoretical findings and the 3D simulations, which is a very sophisticated process, is out of the scope of this work. In Fig.~\ref{fig1}(a), we present a nano dimer with two gold spheres of radii 90 nm and 55 nm, whose linear response is depicted in Fig.~\ref{fig1}(b). The dimer supports two plasmon modes at $\Lambda$=$c/\Omega$=530 nm and $\Lambda_{\rm R}$=$c/\Omega_{\rm R}$=780 nm. $\hat{a}$-mode is driven by a strong laser of wavelength $\lambda_{\rm L}$=593 nm. We place a Raman reporter molecule with radius of 4 nm~\cite{tan2008nanoengineering} (blue) close to the hot spot of the MNP dimer. For a ``proof-of-principle'' demonstration, we consider a single vibrational mode, $\nu=2600$ cm$^{-1}$~\cite{costa2011resonant}, for the Raman reporter molecule. The Stokes signal appears at $\lambda_{\rm R}$=700 nm and couples to $a_{\rm R}$ mode of the double-resonance scheme~\cite{chu2010double}. We model the auxiliary QE by a Lorentzian dielectric function $\epsilon(\omega)$~\cite{wu2010quantum} of resonance $\lambda_{eg}=c/\omega_{eg}$ and damping rate $\gamma_{eg}$. We compare the Raman intensities with the results where no auxiliary QE is present, i.e. enhancement factor. 

In Fig.~\ref{fig3}(a), we also change the position of the auxiliary QE along the z-axis in order to alter the interaction with the MNP. When distance to the hot spot centre (z), increases, the interaction of the plasmon mode with the auxiliary QE, [$f$ in Fig.~\ref{fig3}(b)] decreases. We observe that enhancement occurs at larger wavelengths for stronger MNP-QE coupling as suggested by the basic model. Furthermore, maximum enhancement in Raman signal takes place around $\lambda_{eg}^*\simeq$834 nm, which is farther apart from the Stokes line $\lambda_{\rm R}=$700 nm. A linear Fano resonance would yield the strongest hot spot enhancement when $\lambda_{eg}\simeq \lambda_{\rm R}$~\cite{Stockman2010}. $\lambda_{eg}^*$ is in this regime both for our simple model and for 3D simulations. On the other hand, when auxiliary QE is positioned much closer to the dimer-centre, enhancement decreases 2 orders of magnitude~\cite{supp}. In this regime, excitations cannot be modelled with the presented treatment: strong hybridization arises~\cite{supp}. This is also observed in SHG process~\cite{tasgin2016fluorescence}. Even though many complications may arise in 3D solutions, e.g. coupling to dark modes in the MNP dimer, the results match ``qualitatively'' with our basic model as a ``proof-of-principle'' demonstration. Retardation effects allow Fano resonances to appear in a narrower band compared to our model, similar to Ref.~\cite{turkpence2014engineering}. We also note that our analytical model does not account for the change of density of states, i.e., Purcell factor. 

It can be seen even in the simple setting of Fig.~\ref{fig1}, a setting which can be optimized by further elaboration, an average of a factor of $10^2$ to $10^3$ further enhancement factor is achieved at a QE distance of $\pm$1 nm from the MNPs. It can be said that the hot spot field intensity within the 4 nm gap between MNPs is not a particularly strong one with respect to achievable hot spot enhancement factors of $10^5$-$10^6$ at which an even larger distance between the QE and the MNPs could produce similar orders of further enhancement.

%Controlled positioning at nanoscale can be realized in various experiments. Au nanoparticles, linked to DNA strands, acts as rulers that measures the average wavelength shift of DNA pairs approximately as 1.24nm~\cite{liu2006nanoplasmonic}. Furthermore, plasmonic gaps of bowtie nanoantennas can be selectively filled with dielectric nanoparticles with high efficiency via electron beam lithography process~\cite{hentschel2016linear}.    
%At room temperature the uncertainty in the position or orientation of a dipole of a QE or the constrained Brownian-like motion of the QE even when strongly attached to the MNP, leads to the fact that the total SERS interaction will consist of contributions from all possible Angstrom scale QE-MNP configurations within typical SERS measurement time scales in the range between $10^{-2}$ to $10^{2}$ s and its effects are inevitably going to be manifested in the measured SERS results. So we claim that actually our new discovered enhancement channel must have unknowingly been in action intertwined together with the field enhancement channel in most of the reported SERS enhancement factors in the literature.

\section{Suppression}
\begin{figure}
\centering
\includegraphics[trim=6.5cm 6cm 2.3cm 5cm, clip, width=0.6\textwidth]{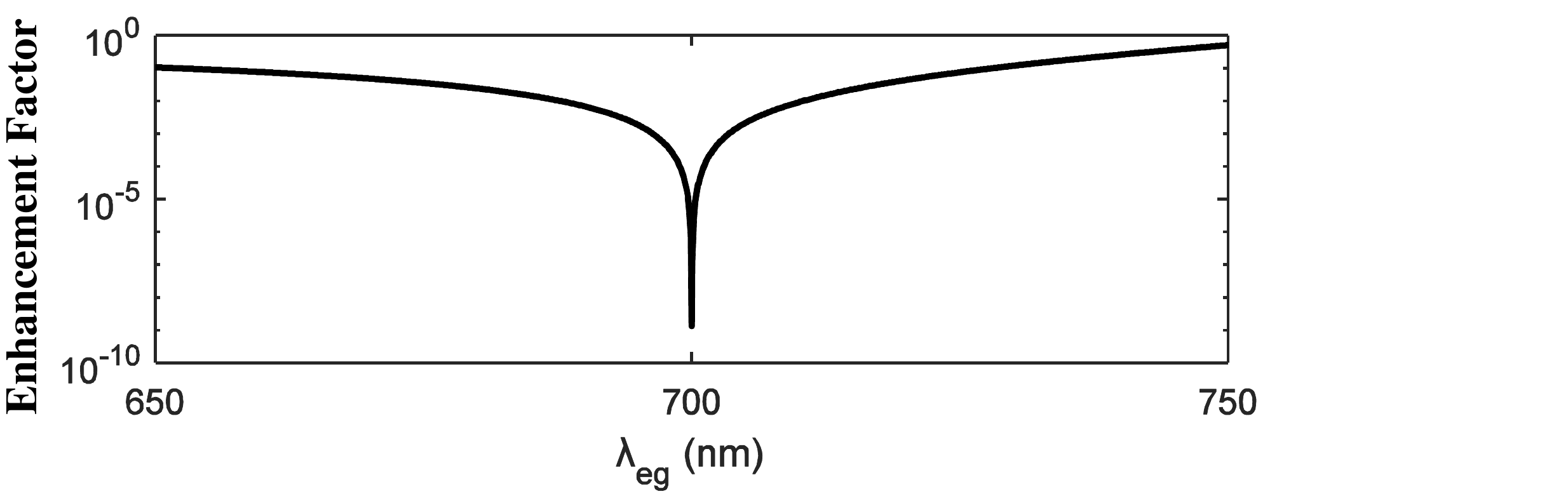}
\caption{Suppression of SERS for $\lambda_{eg}=\lambda_{\rm R}$. When $\omega_{eg}=\omega_{\rm R}$, $\lvert f\rvert^2 y/\gamma_{eg}$ term in the denominator of Eq.~(\ref{R_scattering}) becomes very large, because $\gamma_{eg}$ is very small compared to $\omega_{\rm R}$. This effect can be utilized to avoid losses due to Raman process in fiber laser applications.}  
\label{fig4}
\end{figure}

Our model also predicts that SERS can be suppressed several orders of magnitude, Fig.~\ref{fig4}, for the choice of the auxiliary QE, $\omega_{eg}=\omega_{\rm R}$. Simply, for this case, Fano resonance (transparency) prohibits the plasmon oscillations of the converted frequency $\omega_{\rm R}$ from emerging into the $\hat{a}_{\rm R}$-mode. One can realize that path interference in the nonlinear response is actually not so different from the one taking place in the linear response~\cite{garrido2002classical}. That is, modification of the denominator both in the nonlinear~\cite{turkpence2014engineering, singh2016enhancement} and the linear response~\cite{tacsgin2013metal} have a common form~\cite{tasginlinear2018}.

Similar silencing phenomenon is observed in the SHG experiments and in 3D simulations~\cite{berthelot2012silencing} and can be demonstrated with a simple analytical model~\cite{turkpence2014engineering}. Denominator of Eq.~(\ref{R_scattering}) also shows why a suppression effect can take place similar to the one observed in SHG~\cite{berthelot2012silencing}. If one chooses $\omega_{eg}=\omega_{\rm R}$, the extra term becomes $\gamma_{eg}^{-1}\lvert f\rvert^2 y$. This term is very large since $\gamma_{eg}^{-1}\sim 10^5$ and $f$=0.1 in units scaled with the laser frequency $\omega$ ($\approx$PHz). We stress that Figs.~\ref{fig3}(b) and~\ref{fig4} are generated through the exact time evolutions of Eqs.~(\ref{alpha_R})-(\ref{Rho_ee}). That is, no approximation is used to obtain the results.

On the other hand, the suppression phenomenon --neither in the SHG~\cite{turkpence2014engineering} nor in the Raman cases-- cannot be demonstrated with the 3D simulations of Ref.~\cite{SHG}. This is simply because 3D simulation method~\cite{SHG} is only a first-order approach. Demonstration of the suppression phenomenon necessitates the self-consistent solution of Maxwell equations, as in Eqs.~(\ref{alpha_R})-(\ref{Rho_ee}). Self-consistent 3D simulation of a Raman process is a numerical art on its own.

\section{Summary and Discussions}

 We introduce a new method which can increase the SERS signal without increasing the hot spot intensities. In other words: SERS signal can be further multiplied by a factor of $10^2$-$10^3$, on top of the hot spot formation by plasmon mediated field enhancement, without heating the Raman reporter molecule further. This is different than linear Fano resonances which enhance the hot spot field~\cite{ye2012plasmonic, Zhang2014, he2016near}. The phenomenon takes place due to the modification of the Raman conversion paths, in the presence of an auxiliary QE. Both the $10^2$-$10^3$ enhancement and the unvarying hot spot intensities are confirmed with 3D simulations.   

%further multiplied by a factor of 10^2-10^3, on top of the hot spot formation by plasmon mediated field enhancement (localization ne?).

This phenomenon can be used not only to increase the Raman signal in materials already operating in the break-down or tunnelling regimes and to avoid the modifications of vibrational modes. But it can also be adopted for high spatial resolution imaging of molecules. Raman signal emerges from the region where the two plasmon modes overlap spatially. When this overlap area is kept small, better spatial resolution can be obtained. However, SERS process also weakens with reduced overlap integral. The suggested method can help in increasing the SERS signal to observable values again.

The presented method is not physically intriguing only, but the model provides simple implementations, new phenomena and utilization of new enhancement tools. 

\acknowledgments

MET acknowledges support from TUBITAK Grant No: 1001-117F118 and TUBA-GEBIP 2017 support.

\end{document}